\RequirePackage{fix-cm}
\documentclass[twocolumn]{svjour3}          % twocolumn
\smartqed  % flush right qed marks, e.g. at end of proof
\usepackage{graphicx}
\usepackage{amsmath}
\usepackage{hyperref}
\journalname{Cognitive Neurodynamics}

\begin{document}

\title{%
Sleep, Neuroengineering and Dynamics
}
%\subtitle{}


\author{Jens Christian Claussen \and
Ulrich G.\ Hofmann
}


\institute{%
	    Jens Christian Claussen \at
	    University of Luebeck \\
	    Institute for Neuro- and Bioinformatics\\
	    D-23538 L\"ubeck \\
	    \email{claussen@inb.uni-luebeck.de}\\
\and
Ulrich Hofmann
\at
Neuro-Electronic Systems\\
University Clinic Freiburg \\
D-79095 Freiburg
\\
            \email{%
ulrich.hofmann@uniklinik-freiburg.de
}
}

\date{Cognitive Neurodynamics 6 (3), 211-214 (2012) \\ 
DOI: \href{http://dx.doi.org/10.1007/s11571-012-9204-2}{10.1007/s11571-012-9204-2}
\\Received: 23.04.2012}
%\date{Received: 23.04.2012 }
%\date{Received: 01.08.2011 / Accepted: date}

\maketitle

\mbox{}\vspace*{-4ex}
%\mbox{} \hfill
\noindent
\centerline{\fbox{
\href{http://dx.doi.org/10.1007/s11571-012-9204-2}{\bf Cognitive
Neurodynamics 6 (3), 211-214 (2012)} 
}\mbox{~~~~~~}}%fbox
\\[4ex]

\begin{abstract}
Modeling of consciousness-related phenomena and 
neuroengineering are fields that are rapidly growing together.
We review recent approaches and developments
and point out some promising directions of future research:
 Understanding the dynamics of consciousness states and
associated oscillations, pathological oscillations
as well as their treatment by stimulation, neuroprosthetics
and brain-computer-interface approaches,
and stimulation approaches that probe, influence and
strengthen memory consolidation.
In all these fields, computational models 
connect theory, neurophysiology and neuroengineering research
and pave a way
towards medical applications.
\end{abstract}
Retweeting 
John von Neumann's words
from the 1950s,
``All stable processes we shall predict. All unstable processes we shall
control.'',
it becomes immediate
how
computational neuroscience
 might form
a basis for
novel engineering approaches in neural medicine.
Brain modeling and neural engineering 
are, in footsteps of these words,
aiming at understanding the brain and its emerging
phenomena
on the scientific side,
and interacting with the brain
-- also with the perspective of medical treatment --
from the engineering science side.
Both sides of the theory-experiment coin have always been connected
as, e.g., electrodes are used for data acquisition as well as 
to influence neural systems, be it only by 
well-defined input pulses of a measurement protocol
\cite{stdp}, \cite{ltpltd}.

This roadmap, however, persistently  is a difficult enterprise for various reasons.
The brain is a highly nonlinear dynamical system
and remains a challenge for 
data analysis, theoretical modeling, 
and large-scale computation
\cite{singer},
\cite{kantzschreiber},
\cite{sync},
\cite{chaosinbrain},
\\
\cite{olbrichclaussenachermann}.

In review of these challenges, we would like to emphasize on
four areas of research that mark out outstanding future potential,
and are related to
central keywords:
{\sl Consciousness, Pathological Oscillations, Neuroprosthetics,
and Neuroenhancement}.
In these four areas, computational models in a similar
way connect natural sciences and neuroengineering research
and pave a way
towards medical applications.
\\[2ex]

{\sl Understanding consciousness, anaesthesia and sleep.~--}
Large-scale synchronized oscillations are easily observed 
experimentally and can be both of physiological and pathological character.
The mammal sleep-wake cycle is a remarkably robust oscillation,
in whose regulation again various neural oscillations are involved,
including the cortical slow oscillation in the so-called delta band, 
with frequencies 
around 1 Hz 
\cite{compte03},
\\
\cite{ngo},
\cite{SImattia_sanchezvives}.
\\
Electrical stimulation of the brain at this slow wave frequency
enhanced the slow oscillations themselves as well 
asincreased their memory consolidation effect
\\
\cite{marshall}. 
The slow  oscillations, comprised by the interplay between 
bursting activity and an activity-dependent self-inhibition, 
exhibit a specific anticorrelation in the durations of
Up and Down states; as
Mattia and Sanchez-Vives
show
in comparison of ferret brain slice data, mean-field models,
and simulations
\\
 \cite{SImattia_sanchezvives}.
Mean-field models 
describe the gross activity of a neural subpopulation
(at column level or below) but keep track of
main types of neurons and their simplified connectivity.
They are of great advantage in describing the
consciousness transitions of sleep and general anaesthesia
\cite{SIhutt,SIsteynross}.
As modeled by
% Hutt 
\cite{SIhutt}, GABAergic tonic inhibition
influences the brain's arousal system 
during general anaesthesia including a loss of consciousness.
Including the effect of gap junctions on the dynamics
elucidated the influence on 
 the propensity of generalized seizures
\cite{SIsteynross}.
\\[1ex]

{\sl Medical treatment: pathological oscillations.~--}
It is only a recent development that medicine has spotted the
essential importance of dynamical phenomena 
for understanding and treatment of certain diseases.
These are primarily those where 
oscillations themselves comprise 
the 
disease as in movement disorders, namely essential tremor and
Parkinson's disease,
for which 
electrical stimulation methods co-developed with
theoretical work and computer simulations have found 
their way into clinical practice
\cite{tass}.
But even when there is no observable mechanical or
electrical oscillation,
as in the cortical spreading depression
which is comprised by slow (10$^2$--10$^3$s timescale) Ca$^{2+}$ waves,
control methods may become means of treatment 
\\
\cite{dahlem}.
In epilepsy, and in more severe cases of mood disorders,
deep brain stimulation is applied 
%\\
\cite{abelson}
and sophisticated technical implementations as
radio stimulation
are developed
\\
\cite{delgado}.
Also, mechanical damages to neural pathways can
result in pathological dynamical phenomena,
as in the case of spinal cord injury
which leads to a hyperexcitability of motorneurons.
The modeling approach by
%Venugopal et al.\ 
\cite{SIranu} 
goes down to the role of each ion channels and
aims at a reduction of spasticity.
Finding suitable parameter ranges for
electrical stimulation is annother important issue.
%Krishnamurthi et al.\ 
\\
\cite{SIabbas} 
investigate by measuring velocity reduction
how an optimal amplitude of DBS can be found for Parkinson's disease.
Considering a Rempe-Terman based computational model
of basal ganglia, in
% Njap et al.\ 
\cite{SInjap} 
it is demonstrated that a high concentration of
the inhibiting neurotransmitter GABA together
with electrical stimulation reestablishes
faithful thalamocortical relaying.
A more general question is addressed in
%Sch\"utt et al.\ 
%\\
\cite{SIschuett} 
by investigating low- to high-frequency stimulation in an
Izhikevich-type cortical network model,
with the observation that in a frequency range
around 100Hz, a dynamical desynchronization is observed.
\\[1ex]

{\sl Neuroprosthetics and Brain-Computer Interfaces.~--}
The consequent continuation of 
few-electrode stimulations and recordings are
Human-Machine interfaces
(HMIs)
\cite{birbaumer},
the most apparent applications where 
theoretical brain science and engineering meet.
In the loss of direct control through the natural pathways, prosthetic
devices have to be controlled through 
an HMI
which comprises a 
``thought-control''
\cite{hochberg,pfurtscheller}.
The consequent continuation of that idea -- and most immediate demonstration
of an HMI
at work 
is
that a blind patient can use the HMI for reading 
\cite{zrenner},
cochlea implants 
improve hearing
\cite{edgerton},
and patients with locked-in syndrome can use 
an EEG-based
BCI for
expressing words  
\cite{wolpaw}.
Overall, prosthetic applications require a sound modeling 
approach and understanding of the neural
processing, and, if possible, also coding, 
to design an effective information
interfacing with
the brain.
The visual system is a part of the brain where
experimental research and detailed modeling 
have made large progress.
Here,
%Norheim et al.\ 
\cite{SInorheim_einevoll} 
and 
%Einevoll et al.\ 
\cite{SIeinevollplesser} 
\\
investigate both a minimal and a feedback - extended model for 
temporal processing in the lateral geniculate nucleus (LGN).
The final step of brain-machine interfacing 
goes towards dissolution of the border
between computer and brain itself:
In their beautiful experimental setup,
% Perez-Marcos, Sanchez-Vives and Slater 
%\\
\cite{SIsanchezvives2} 
demonstrate virtual hand illusions and 
 -- at least for a part of the body --
question the conscious awareness of our Self,
and thereby 
connect
two quite different fields.
\\[1ex]

{\sl Understanding and influencing neural plasticity.~--}
Neural plasticity, the basis of all learning, 
can not only be influenced pharmacologically, but 
also by various means of electrical stimulation with
remarkable effects on cognitive learning and consolidation
\\
\cite{marshall}.
Stimulation, as well as learning, can effect on both  
the dynamics 
\cite{SImattia_sanchezvives},
\cite{SIschuett} 
and on the plasticity 
\cite{SIclopath},
\cite{SIvogt}.
%In 
\cite{SIvogt} 
%Vogt et al.\
demonstrate modulatory effects of dopamine
based on an underlying STDP learning. 
Modulatory or multi-input based learning mechanisms 
are good candidates to explain memory consolidation. 
In this direction,
% Clopath
 \cite{SIclopath} 
compares two recently proposed mechanisms,
namely tag-trigger consolidation, and
a metastate tagging model,
and provides a comprehensive comparison of both concepts.
Finally,
% Weigenand et al.\ 
\cite{weigenand11},
\cite{SIweigenand} 
investigate the phase-dependence of stimulation
of cortical slow waves.
It would be highly interesting,
based on 
improved
 understanding of neural coding and plasticity
mechanisms,
to 
design specific stimulation protocols that
selectively strengthen desired acquired memories.
\\[1ex]
% \clearpage

{\sl Outlook.}~~
While it is tempting to go beyond
the purpose of medical treatment 
by  ``enhancing the brain''
\cite{farah},
a dee\-per understanding of neural plasticity mechanisms
by theoretical and computational models also is expected to offer
pathways to the treatment of various neural disabilities and disorders,
be them memory-related like 
Alzheimer, 
or mood-related disorders as depressive disorders.
How emotion modulates learning, and how emotional processes
dynamically regulate psychological sta\-tes
is an emerging field
\cite{helland2008,huber_braun1999}
and can be expected to become the 
`neuroengineering''
extension of 
of computational modeling
for the treatment of brain disorders.

{\small
{\sl Acknowledgments.~--}
The collection of articles 
on these topics in the current and the subsequent 
issue
of {\sl Cognitive Neurodynamics}
emerged from the 1$^{\rm st}$ Baltic Autumn School
``Applied Computational Neuroscience:
Sleep, Neuroengineering and Dynamics'',
hosted by the University of L\"ubeck 
with generous support from the 
Deut\-sche Forschungsgemeinschaft (DFG).
We thank all invitees and participants for their
contributions for this Special Issue
as well as for their
enthusiastic and stimulating discussions 
during the workshop.
}%small

%\clearpage

%\bibliographystyle{spphys} 
%\bibliographystyle{apalike}%doesnotwork 

\end{document}